\documentclass[10pt,english]{IEEEtran}
\usepackage[T1]{fontenc}
\usepackage{geometry}
\usepackage{amsmath}
\usepackage{xpatch}
\usepackage{amssymb}
\usepackage{esint}
\usepackage{mathtools}
\usepackage{amsthm}
\usepackage{nicefrac}
\usepackage{gensymb}
\usepackage{bigints}
\usepackage{mathtools}
\usepackage{blkarray, bigstrut}
\usepackage{physics}
\usepackage{calligra}
\usepackage{graphicx}
\usepackage{epstopdf}
\usepackage{dsfont}
\usepackage{array,ragged2e}
\usepackage{enumitem}
\usepackage{etoolbox}
\usepackage{babel}
\usepackage[nopar]{lipsum}
\usepackage{psfrag}
\usepackage[ruled,norelsize] {algorithm2e}
\usepackage{balance}
\usepackage{float}
\usepackage{verbatim}
\usepackage[caption = false]{subfig}
\SetKw{KwBy}{by}
\usepackage{algpseudocode}
\usepackage{makecell}
\usepackage{to-be-determined} 

\usepackage[compact]{titlesec}

\titlespacing{\section}{0.5pt}{*0.5}{*0.5}
\titlespacing{\subsection}{0.5pt}{*0.5}{*0.5}
\titlespacing{\subsubsection}{0.5pt}{*0.5}{*0.5}

\makeatletter
\newcommand{\removelatexerror}{\let\@latex@error\@gobble}
\makeatother

\newtheoremstyle{plain}
  {\topsep}   
  {\topsep}   
  {\itshape}  
  {0pt}       
  {\bfseries} 
  {.}         
  {5pt plus 1pt minus 1pt} 
  {\thmname{#1}\thmnumber{ #2} \textnormal{(\thmnote{#3})}} 

\xpatchcmd{\proof}{\hskip\labelsep}{\hskip5\labelsep}{}{}  
\makeatletter
\xpatchcmd{\proof}{\@addpunct{.}}{\@addpunct{:}}{}{}
\makeatother

\renewcommand\[{\begin{equation}}
\renewcommand\]{\end{equation}} 
\usepackage{listings}
\usepackage{fancyvrb}
\usepackage{framed}

\usepackage{courier}
\definecolor{dkgreen}{rgb}{0,0.3,0}
\definecolor{gray}{rgb}{0.5,0.5,0.5}

\begin{document}

\title{Joint Explainability and Sensitivity-Aware Federated Deep Learning for Transparent 6G RAN Slicing}
\author{Swastika Roy$^{(1,2)}$, Farhad Rezazadeh$^{(1,2)}$, Hatim Chergui$^{(3)}$, and Christos Verikoukis$^{(4,5)}$\\
{\normalsize{} $^{(1)}$ Telecommunications Technological Center of Catalonia (CTTC), Barcelona, Spain}\\
{\normalsize{} $^{(2)}$ Technical University of Catalonia (UPC), Barcelona, Spain}\\
{\normalsize{} $^{(3)}$ i2CAT Foundation, Barcelona, Spain}\\
{\normalsize{} $^{(4)}$ University of Patras, Greece}\\
{\normalsize{} $^{(5)}$ ISI/ATHENA, Greece}\\
{\normalsize{}Contact Emails: \texttt{\{sroy, frezazadeh\}@cttc.es, chergui@ieee.org, cveri@ceid.upatras.gr}}}

\maketitle
\thispagestyle{empty}

\begin{abstract}
In recent years, wireless networks are evolving complex, which upsurges the use of zero-touch artificial intelligence (AI)-driven network automation within the telecommunication industry. In particular, network slicing, the most promising technology beyond 5G, would embrace AI models to manage the complex communication network. Besides, it is also essential to build the trustworthiness of the AI black boxes in actual deployment when AI makes complex resource management and anomaly detection. Inspired by closed-loop automation and Explainable Artificial intelligence (XAI), we design an Explainable Federated deep learning (FDL) model to predict per-slice RAN dropped traffic probability while jointly considering the sensitivity and explainability-aware metrics as constraints in such non-IID setup. In precise, we quantitatively validate the faithfulness of the explanations via the so-called attribution-based \emph{log-odds metric} that is included as a constraint in the run-time FL optimization task. Simulation results confirm its superiority over an unconstrained integrated-gradient (IG) \emph{post-hoc} FDL baseline.
\end{abstract}
\vspace{-4mm}
\begin{IEEEkeywords}
6G, classification, FL , game theory, proxy-Lagrangian, SLA, stochastic policy, traffic drop, XAI, ZSM
\end{IEEEkeywords}

\section{Introduction}
The most promising  6G network slicing technology insists on adopting autonomous management and orchestration of the end-to-end (E2E) network resources at the network domains because the isolation of slices may induce a high cost in terms of efficiency \cite{2,3}.
\begin{figure}[h]
\centering
    \includegraphics[angle=360,width=90mm, scale=0.37]{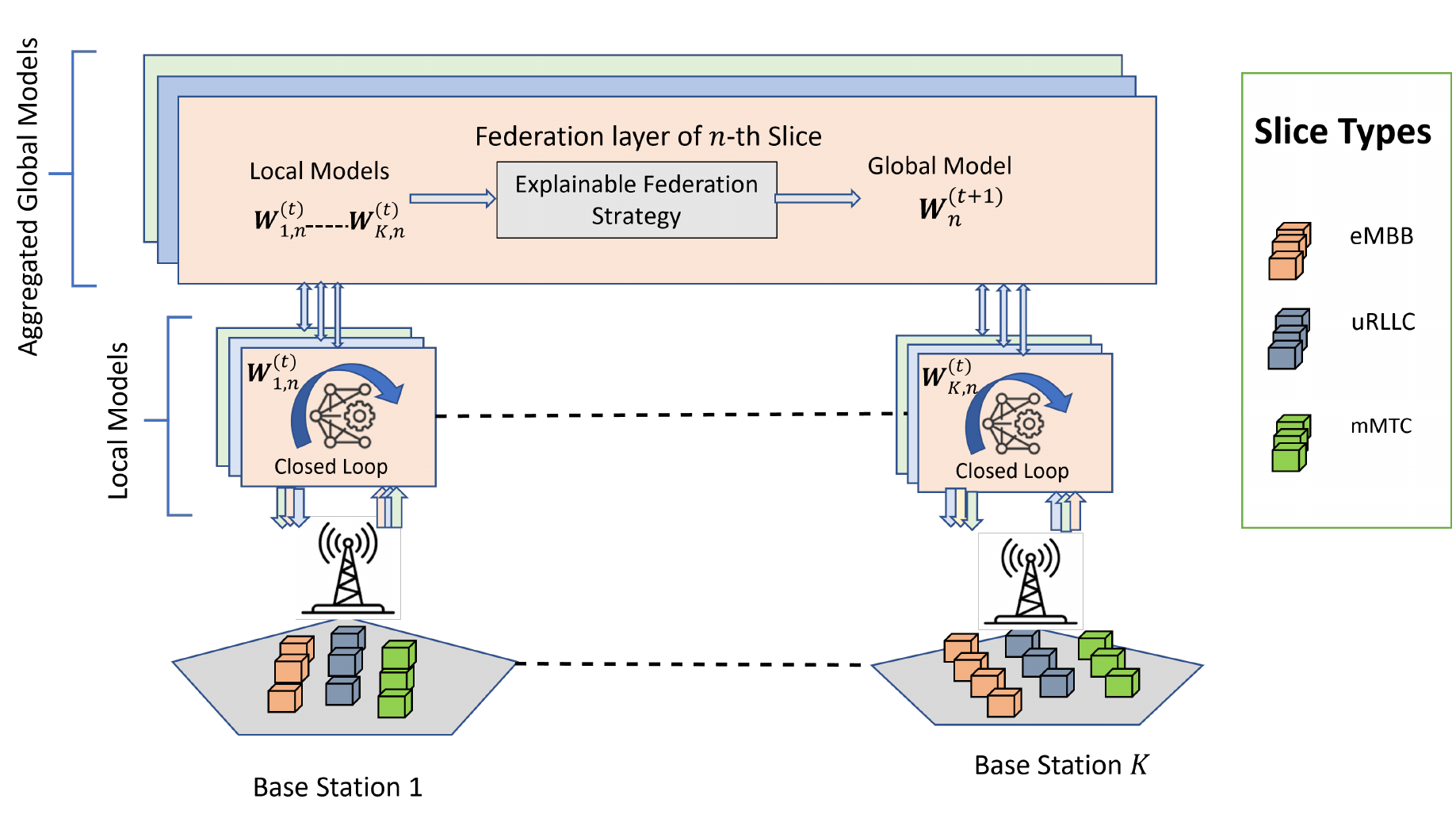}
\caption{RAN federated traffic drop classification in NS}
\vspace{-7mm}
\end{figure}
So, ETSI standardized zero-touch network and service management (ZSM) framework has been considered \cite{ZSM}.Here, zero-touch refers to the automation and management of resources without human interference.
Besides, developing cognitive slice management solutions in 6G networks is essential to automatically orchestrate and manage network slices, particularly network resources across different technological domains (TDs), along with ensuring the end-user's QoE and QoS \cite{a,b}. Hence, the \cite{AI} has proposed an AI-native network slicing management solution of 6G networks to support emerging AI services. 
Also, AI algorithms should be driven by the distributed nature of datasets to acquire the full potential of network slicing automation, which will solve the problematic behavior of the cloud-centric traditional ML schemes. Thus, a decentralized learning approach is required to handle distributed network slices efficiently. 
For this, we choose Federated learning (FL) \cite{FL,frah} to handle distributed network slices efficiently like our another research work \cite{mine}.
Besides, even if DNN hold the state-of-the-art \cite{ns1,ns2,ns3} in solving resource allocation and orchestration problems of network slicing, the black-box nature of such ML models impedes understanding of their decisions, any flaws in the datasets or the model's performance behavior. Moreover, the 6G network is going to be "machine-centric" technology which signifies that all the corresponding "smart things" in the 6G network will operate intelligently but as a smart black box \cite{XAI2}. Here, the smart black box is not transparent in its action or decision-making processes and could have adverse effects on the network's operations of the 6G technology. In this concern, XAI provides human interpretable methods to adequately explain the AI system and its decisions for gaining the human's trust in the loop.
Also, \cite{XAI} indicates that it is a prerequisite of any ZSM-based AI models in 6G to enrich translucency of their models. Viewing this fact, zero-touch XAI-driven FL will be fetching a particular emphasis for its automation and unique advantages, which are essential for end-user trust and secured procedure. In contrast, the conventional XAI focuses only on the interpretability and transparency of any ML system.
Some works of XAI  \cite{ex1,ex2, know} indicate the importance of explainability and present some research works on handover and resource allocation, etc., in the beyond 5G networks. In \cite{phy}, XAI for physical/MAC layers in 6G networks are focused. In comparison, the authors of \cite{XAI_FL} present a trust-aware federated deep reinforcement learning-based device selection technique in an autonomous driving scenario.
And, to evaluate the performance of XAI models, the paper \cite{xai_metric} introduces some essential metrics. So, in this work, we will present a novel zero-touch Explainable Federated learning (FL) as the decentralized approach for traffic drop classification in 6G network slices \cite{FL}.

\subsection{Contributions}
In this paper, we present the following contributions 
\begin{itemize}

    \item We introduce a novel iterative explainable federated learning approach, where a constrained traffic drop detection classifier and an \emph{explainer} exchange---in a closed loop way--- attributions of the features as well as predictions to achieve a transparent zero-touch service management of 6G network slices at RAN in a non-IID setup.
    
    \item We adopt the integrated gradients XAI method to showcase features attributions.
    
    \item The generated attributions are then used to quantitatively validate the faithfulness of the explanations via the so-called \emph{log-odds} metric which is included as a constraint in the FL optimization task.
    
    \item We formulate the corresponding joint recall and log-odds-constrained FL optimization problem under the \emph{proxy-Lagrangian} framework and solve it via a non-zero sum two-player game strategy \cite{{TwoPlayer}}, while comparing with the unconstrained integrated-gradient post-hoc FL baseline.    
\end{itemize}

\section{RAN Architecture and Datasets}
A shown in Fig. 1, we consider a radio access network (RAN), which is composed of a set of $K$ the base station (BSs), wherein a set of $N$ parallel slices are deployed. Each BS runs a local control closed-loop (CL) which collects monitoring data and performs traffic drop prediction. Specifically, the collected data serves to build local datasets for slice $n \,(n=1,\ldots,N)$, i.e., $\mathcal{D}_{k,n}=\{\mathbf{x}_{k,n}^{(i)},y_{k,n}^{(i)}\}_{i=1}^{D_{k,n}}$, where $\mathbf{x}_{k,n}^{(i)}$ stands for the input features vector while $y_{k,n}^{(i)}$ represents the corresponding output. In this respect, Table I summarizes the features and the output of the local datasets. These accumulated datasets are non-IID due to the different traffic profiles induced by the heterogeneous users' distribution and channel conditions. Moreover, since the collected datasets are generally non-exhaustive to train accurate anomaly detection classifiers, the local CLs take part in a federated learning task wherein an E2E slice-level federation layer plays the role of a model aggregator.
\begin{table}[t!]
\label{Datasets-tab}
\centering	

\caption{Dataset Features and Output}
{\color{black}\begin{tabular}{lc}
\hline
\hline 

Feature & Description\\
\hline
\texttt{Average PRB}& Average Physical Resource Block \\ 
\texttt{Latency} & Average transmission latency\\
\texttt{Channed Quality} &  SNR value expressing the wireless channel quality\\

\hline 
\hline 
& \\
\hline
\hline 
Output & Description\\
\hline
\texttt{Dropped Traffic} & Probability of dropped traffic(\%)\\
\hline
\hline
\label{Datasets-tab1}
\end{tabular}
}
\end{table}

\vspace{-1mm}
\section{Explainable FDL for Transparent Traffic Drop Classification}
Here, we describe the different stages of the joint explainability and sensitivity-aware FDL as summarized in Fig. 2.
\subsection{Closed-Loop Description}
We propose a federated deep learning architecture where the local learning is performed iteratively with run-time explanation in a closed loop way as shown in Fig. 2. We design a deep neural network FL model. For each local epoch, the Learner module feeds the posterior symbolic model graph to the Tester block which yields the test features and the corresponding predictions $\hat{y}_{k,n}^{(i)}$ to the Explainer. The latter first generates the features attributions using  integrated gradients XAI method. The \emph{Log-odds Mapper} then uses these attributions to select the top $p$ features that are then masked. The corresponding soft probability outputs are afterward used to calculate the the log-odds (LO) metric that is fed back to the Learner to include it in the local constrained optimization in step 6. Similarly, the \emph{Recall Mapper} calculate the recall score $\rho_{k,n}$ based on the predicated and true positive values at stage 3 and 4 to include it in the local constrained optimization in step 6.
Indeed, for each local CL $(k,n)$, the predicted traffic drop class $\hat{y}_{k,n}^{(i)},\,(i=1,\ldots,D_{k,n})$, should minimize the main loss function with respect to the ground truth $y_{k,n}^{(i)}$, while jointly respecting some long-term statistical constraints defined over its $D_{k,n}$ samples and jointly corresponding to recall and explainability log-odds. 

As shown in steps 1 and 7 of Fig. 2, the optimized local weights at round $t$, $\mathbf{W}_{k,n}^{(t)}$, are sent to the server which generates a global FL model for slice $n$ as,
\begin{equation}
\label{GlobalFL}
    \mathbf{W}_{n}^{(t+1)}=\sum_{k=1}^K \frac{D_{k,n}}{D_n}\mathbf{W}_{k,n}^{(t)},
\end{equation}
where $D_n=\sum_{k=1}^K D_{k,n}$ is the total data samples of all datasets related to slice $n$. The server then broadcasts the global model to all the $K$ CLs that use it to start the next round of iterative local optimization. Specifically, it leverages a two-player game strategy to jointly optimize over the objective  and original constraints as well as their smoothed surrogates and detailed in the sequel.

\begin{figure}[t!]
\centering
    \includegraphics[scale=0.55,trim={2.2cm 0 0 0},clip]{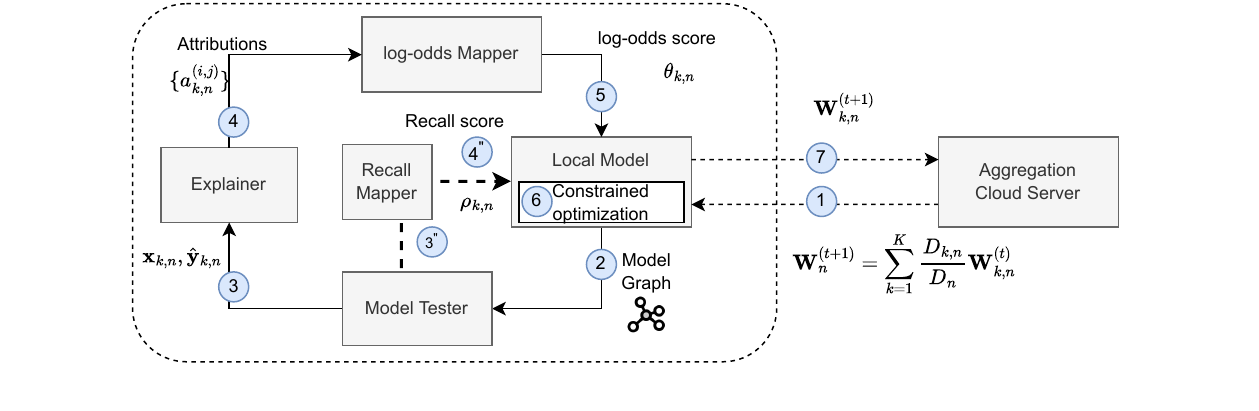}
\vspace{-4mm}    
\caption{Explainable FDL building blocks}
\label{turbofl}
\vspace{-6mm}
\end{figure}

\subsection{Model Testing and Explanation}

As depicted in stage 2 of Fig. 2, upon the reception of the updated model graph, the Tester uses a batch drawn from the local dataset to reconstruct the test predictions $\hat{\mathbf{y}}_{k,n}^{(i)}$. All the graph, test dataset and the predictions are fed to the Explainer at stage 3. After that, at stage 4, Explainer generates the attributions
by leveraging the low-complexity Integrated Gradient (IG) scheme \cite{IG}, which is based on the gradient variation when sampling the neighborhood of a feature.
Attributions are a quantified impact of each single feature on the predicted output. Let $\mathbf{a}_{k,n}^{(i)}\in \mathbb{R}^{Q}$ denote the attribution vector of sample $i$, which can be generated by any attribution-based XAI method.

\subsection{Log-odds Mapping}

To characterize the trustworthiness of the local model, we calculate the log-odds metric, $\theta_{k,n}$ \cite{log}. It measures the influence of the top-attributed features on the model's prediction. Specifically, the log-odds score is defined as the average difference of the negative logarithmic probabilities on the predicted class before and after masking the top $p$\% features with zero padding \cite{log}.
In this respect, the \emph{log-odds Mapper} at stage 5 of Fig. 2 starts by selecting top $p$\% features based on their attributions which is collected from stage 4 and replace them with zero padding. That is,
\begin{equation}
\label{log_odds}
    \theta_{k,n} = - \frac{1}{D_{k,n}}\sum_{i=1}^{D_{k,n}} \log \frac{\Pr\left(\hat{y}_{k,n}^{(i)}|\hat{\mathbf{x}}_{k,n}^{(i)}\right)}{\Pr\left(\hat{y}_{k,n}^{(i)}|\mathbf{x}_{k,n}^{(i)}\right)}, 
\end{equation}
where, $\hat{y}_{k,n}^{(i)}$ is the predicted class, $\mathbf{x}_{k,n}^{(i)}$ are the features in the original dataset and $\hat{\mathbf{x}}_{k,n}^{(i)}$ denotes the features in the modified dataset with top $p$\% features zero-padded. Finally, the log-odds Mapper reports the log-odds score, which is used as one of the constraints for the constrained FL optimization task.

\subsection{Joint Recall and Explainability-Aware Traffic Drop Classification}
Besides the log-odds score used for explainability,as shown in steps 3 and 4, we invoke the \emph{recall} as a measure of the sensitivity of the FL local classifier, which we denote $\rho_{k,n}$, i.e.,
\label{recall}
\begin{equation}
    \rho_{k,n}= \pi^{+}\left(\mathcal{D}_{k,n}\left[\hat{y}_{k,n}^{(i)}=1\right]\right)
\end{equation}
Where, $\pi^{+}(\mathcal{D}_{k,n})$ defines the proportion of $\mathcal{D}_{k,n}$ classified positive, and $\mathcal{D}_{k,n}[*]$ is the subset of $\mathcal{D}_{k,n}$ satisfying expression *.

In order to trust the traffic drop anomaly detection/classification, a set of AI SLA is established between the slice tenant and the infrastructure provider, where a lower bound $\alpha_{n}$ is imposed to the recall score, while an upper bound $\beta_{n}$ is set for the log-odds score.
This translates into solving a constrained local classification problem in iterations specified by the epochs as well as in FL rounds $t\,(t=0,\ldots,T-1)$ i.e.,

\begin{subequations}
\label{OPT1}
\begin{equation}
    \min_{\mathbf{W}_{k,n}^{(t)}}\, \frac{1}{D_{k,n}}\sum_{i=1}^{D_{k,n}}\ell\left(y_{k,n}^{(i)}, \hat{y}_{k,n}^{(i)}\left(\mathbf{W}_{k,n}^{(t)},\mathbf{x}_{k,n}\right)\right),
\end{equation}

\begin{equation}
     \mathrm{s.t.}\hspace{5mm}\rho_{k,n} \geq \alpha_{n}\label{recall},
\end{equation}
\begin{equation}
     \hspace{7mm}\theta_{k,n} \leq \beta_{n}\label{log},
\end{equation}
\end{subequations}
which is solved by invoking the so-called \emph{proxy Lagrangian} framework \cite{Cotter}, since the recall is not a smooth constraint. This consists first on constructing two Lagrangians as follows:
\begin{subequations}
\label{ProxyLagrangian}
\begin{equation}
\begin{split}
    \mathcal{L}_{\mathbf{W}_{k,n}^{(t)}}=&\frac{1}{D_{k,n}}\sum_{i=1}^{D_{k,n}}\ell\left(y_{k,n}^{(i)}, \hat{y}_{k,n}^{(i)}\left(\mathbf{W}_{k,n}^{(t)},\mathbf{x}_{k,n}\right)\right)
    \\& +\lambda_1\Psi_1\left(\mathbf{W}_{k,n}^{(t)}\right)+\lambda_2\Psi_2\left(\mathbf{W}_{k,n}^{(t)} \right),
\end{split}
\end{equation}
\begin{equation}
    \mathcal{L}_{\lambda}=\lambda_1\Phi_1\left(\mathbf{W}_{k,n}^{(t)}\right)+\lambda_2\Phi_2\left(\mathbf{W}_{k,n}^{(t)}\right)
\end{equation}
\end{subequations}
where $\Phi_{1,2}$ and $\Psi_{1,2}$ represent the original constraints and their smooth surrogates, respectively. In this respect, the recall surrogate is given by,
\begin{equation}
    \Psi_1 = \frac{\sum_{i=1}^{D_{k,n}}{y}_{k,n}^{(i)} \times \min\Bigl\{\hat{y}_{k,n}^{(i)}, 1\Bigl\}}{\sum_{i=1}^{D_{k,n}} {y}_{k,n}^{(i)}} - \alpha_n
\end{equation}
while $\Psi_2=\Phi_2= \beta_n - \theta_{k,n}$ since the negative logarithm is already a convex function. It also confirms that the solutions of the optimization problem are equivalent to those obtained if only the original constraints were used.

This optimization task turns out to be a non-zero-sum two-player game in which the $\mathbf{W}_{k,n}^{(t)}$-player aims at minimizing $\mathcal{L}_{\mathbf{W}_{k,n}^{(t)}}$, while the $\lambda$-player wishes to maximize $\mathcal{L}_{\lambda}$ \cite[Lemma 8]{TwoPlayer}. While optimizing the first Lagrangian w.r.t. $\mathbf{W}_{k,n}$ requires differentiating the constraint functions $\Psi_1(\mathbf{W}_{k,n}^{(t)})$ and $\Psi_2(\mathbf{W}_{k,n}^{(t)})$, to differentiate the second Lagrangian w.r.t. $\lambda$ we only need to evaluate $\Phi_1\left(\mathbf{W}_{k,n}^{(t)}\right)$ and $\Phi_2\left(\mathbf{W}_{k,n}^{(t)}\right)$. Hence, a surrogate is only necessary for the $\mathbf{W}_{k,n}$-player; the $\lambda$-player can continue using the original constraint functions. The local optimization task can be written as,
\begin{subequations}
\label{ProxyLagrangian}
\begin{equation}
\begin{split}
    \min_{\mathbf{W}_{k,n}\in \Delta} \,\,\,\,\max_{\lambda,\, \norm{\lambda}\leq R_\lambda}\,\,\mathcal{L}_{\mathbf{W}_{k,n}^{(t)}}
\end{split}
\end{equation}
\begin{equation}
        \max_{\lambda,\, \norm{\lambda}\leq R_\lambda}\,\,\,\,\min_{\mathbf{W}_{k,n}\in \Delta} \mathcal{L}_{\lambda},
\end{equation}
\end{subequations}

where thanks to Lagrange multipliers, the $\lambda$-player chooses how much to weigh the proxy constraint functions, but does so in such a way as to satisfy the original constraints, and ends up reaching a nearly-optimal nearly-feasible solution \cite{Gordon}. These steps are all summarized in Algorithm 1.

\algblock{ParFor}{EndParFor}
\algnewcommand
\algorithmicparfor{\textbf{parallel for}}
\algnewcommand\algorithmicpardo{\textbf{do}}
\algnewcommand
\algorithmicendparfor{\textbf{end parallel for}}
\algrenewtext{ParFor}[1]{\algorithmicparfor\ #1\ \algorithmicpardo}
\algrenewtext{EndParFor}{\algorithmicendparfor}

\begin{algorithm}[t]
\caption{Explainable Federated Deep Learning}
\footnotesize
\SetAlgoLined
\KwIn{$K$, $m$, $\eta_{\lambda}$, $T$, $L$. \texttt{\# See Table II}}

Server initializes $\mathbf{W}_n^{(0)}$ and broadcasts it to the $\mathrm{CL}$s\\

\For{$t=0,\ldots,T-1$}{
\textbf{parallel for} $k=1,\ldots,K$ \textbf{do}\\
 Initialize $M=$ \texttt{num\_constraints} and $\mathbf{W}_{k,n,0}=\mathbf{W}_{n}^{(t)}$\\ 
 Initialize $\mathbf{A}^{(0)}\in \mathbb{R}^{(M+1) \times (M+1)}$ with $\mathbf{A}_{m',m}^{(0)}=1/(M+1)$\\
 \For {$l=0,\ldots,L-1$}{
 Receive the graph $\mathcal{M}_{k,n}$ from the local model\\
  \texttt{\# Test the local model and calculate the attributions}\\
  $a_{k,n}^{i,j}=$ \texttt{Int. Gradient} $\left(\mathcal{M}_{k,n}\left(\mathbf{W}_{k,n,l}, \mathbf{x}_{k,n}\right)\right)$\\
 
  \texttt{\# Mask the top p\% dataset based on the attributions with zero padding}\\
  \texttt{\# Calculate the log-odds metric}\\
   $\theta_{k,n}$ = $\frac{1}{D_{k,n}}\sum_{i=1}^{D_{k,n}} \log \frac{\Pr\left(\hat{y}_{k,n}^{(i)}|\hat{\mathbf{x}}_{k,n}^{(i)}\right)}{\Pr\left(\hat{y}_{k,n}^{(i)}|\mathbf{x}_{k,n}^{(i)}\right)}$\\

   \texttt{\# Calulate the recall metric}\\
   $\rho_{k,n}= \pi^{+}\left(\mathcal{D}_{k,n}\left[\hat{y}_{k,n}^{(i)}=1\right]\right)$\\
   
  Let $\lambda^{(l)}$ be the top eigenvector of $\mathbf{A}^{(l)}$\\
  \texttt{\# Solve problem (\ref{OPT1}) via oracle optimization}\\
  Let $\hat{\mathbf{W}}_{k,n,l}=\mathcal{O}_\delta\left(\mathcal{L}_{\mathbf{W}_{k,n,l}}(\cdot, \hat{\lambda}^{(l)})\right)$\\
  Let $\Delta_{\lambda}^{(l)}$ be a gradient of $\mathcal{L}_{\lambda}(\hat{\mathbf{W}}_{k,n,l}, \lambda^{(l)})$ w.r.t. $\lambda$\\
  \texttt{\# Exponentiated gradient ascent}\\
  Update $\tilde{\mathbf{A}}^{(l+1)}=\mathbf{A}^{(l)}\odot\cdot\exp{\eta_{\lambda}\Delta_{\lambda}^{(l)}(\lambda^{(l)})}$\\
  \texttt{\# Colunm-wise normalization}\\
  $\mathbf{A}_{m}^{(l+1)}=\tilde{\mathbf{A}}_{m}^{(l+1)}/\norm{\mathbf{A}_{m}^{(l+1)}}_1,\,m=1,\ldots,M+1$\\
  }
  \Return{$\hat{\mathbf{W}}_{k,n}^{(t)}=\frac{1}{L^{\star}}\sum_{l=0}^{L-1}\hat{\mathbf{W}}_{k,n,l}$}\\
Each local CL $(k,n)$ sends $\hat{\mathbf{W}}_{k,n}^{(t)}$ to the server. \\
 \textbf{end parallel for}\\
 \Return{$\mathbf{W}_{n}^{(t+1)}=\sum_{k=1}^K \frac{D_{k,n}}{D_n}\hat{\mathbf{W}}_{k,n}^{(t)}$}\\
 and broadcasts the value to all local CLs.}
\end{algorithm}
  
\vspace{-1mm}

\section{Results}

This section analyzes the proposed Closed loop EFL framework in detail. To build the explainability-aware constrained traffic drop classification model, we use feature attributions which is the pillar of this approach. 
After that, we present the impact of considering jointly the recall and log-odds metrics as constraints for optimizing the FL classification problem by showing results of FL convergence and log-odds score. Finally, we study the correlation between features attributions, observed predictions, and \emph{true} predictions and draw some important conclusions. Specifically, to implement the model Tester and Explainer, we invoke \texttt{DeepExplain} framework, which includes state-of-the-art gradient and perturbation-based attribution methods \cite{Deep}. It provides an attribution score based on the feature's contribution to the model's output, which we integrate with our proposed constrained traffic drop classification FL framework in a closed-loop iterative way.
\vspace{-2mm}
\subsection{Parameter Settings and Baseline}
Three primary slices eMBB, uRLLC and mMTC are considered to analyze the proposed Explainable FL policy. Here, the datasets are collected from the BSs and the overall summary of those datasets are presented in Table II. We use vector $\beta$ for the explainability lower bound threshold and $\alpha$ for the upper bound of recall score corresponding to the different slices. As a baseline, we adopt a vanilla FL \cite{Van_FL} with post-hoc integrated gradient explanation,that is, a posterior explanation performed upon the end of the FL training.
\vspace{-4mm}
\begin{table}[htb]
\label{Table1}
\centering	
\newcolumntype{M}[1]{>{\centering\arraybackslash}m{#1}}
\caption{Settings}
\begin{tabular}{ccc}
\hline 
\hline
Parameter & Description & Value\\
\hline
DNN & Deep neural network size & $2$-hidden layers with $10$ nodes\\
$N$ & \# Slices & $3$\\
$K$ & \# BSs & $50$\\ 
$D_{k,n}$ & Local dataset size & $800$ samples\\ 
$T$ & \# Max FL rounds & $50$\\  
$U$ & \# Total users (All BSs)& $15000$\\
$L$ & \# Local epochs & $100$\\ 
$R_{\lambda}$ & Lagrange multiplier radius &  Constrained: $10^{-5}$\\
$\eta_{\lambda}$ & Learning rate & $0.02$ \\ 
\hline
\hline
\label{FLsettings}
\end{tabular}

\end{table}
\vspace{-8mm}
\subsection{Result Analysis}
In this scenario, resources allocated to slices according to their traffic patterns and radio conditions while ensuring a long term isolation via the constraints.

\begin{figure}[htb]
\centering
\includegraphics[width=0.40\textwidth]{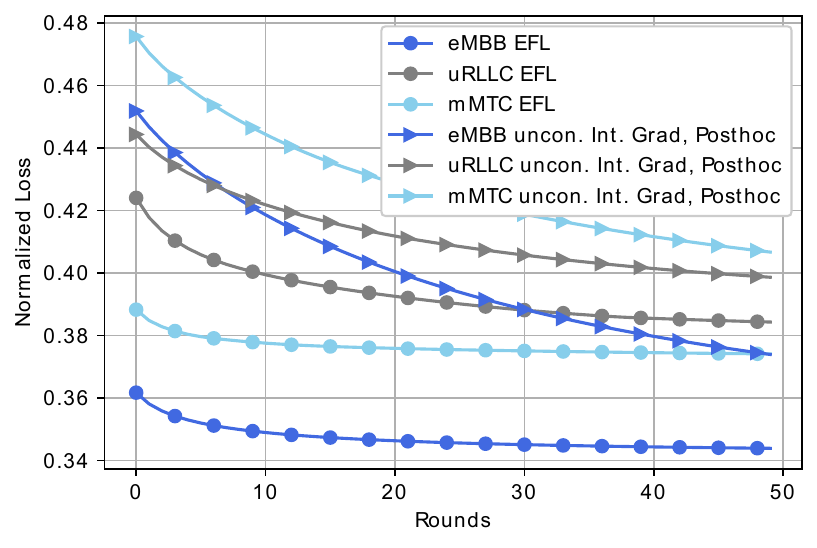}
\caption{ Analysis of FL training loss vs FL rounds of Proposed EFL with Lower bound of Recall score, $\alpha = [0.9, 0.95, 0.95]$ and Upper bound of log-odds score, $\beta = [-0.01, -0.01 , -0.01]$}
\vspace{-7mm}
\end{figure}

\begin{figure}[htb]
\centering
     \includegraphics[width=0.40\textwidth]{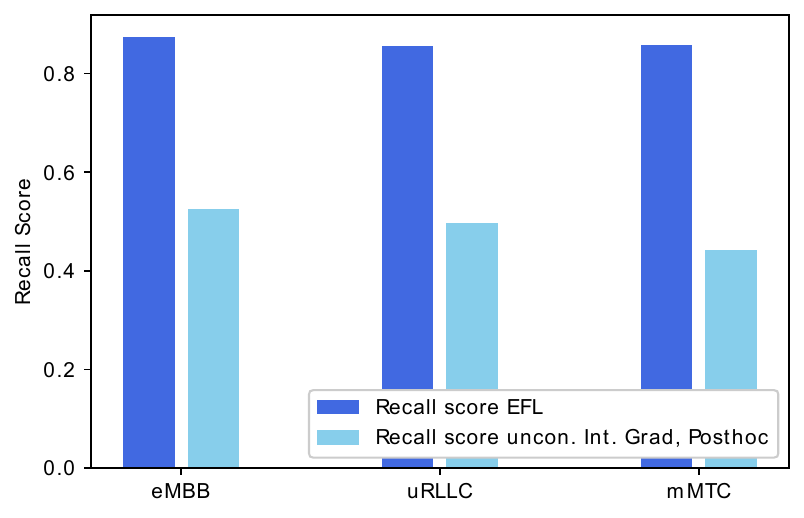}
\caption{ Analysis of Recall score with Lower bound of Recall score, $\alpha = [0.9, 0.95, 0.95]$ and Upper bound of log-odds score, $\beta = [-0.01, -0.01 , -0.01]$}
\end{figure}

\begin{itemize}
\item \textbf{Convergence:} 
As depicted in Fig. 3, we can conclude that the proposed constrained EFL resource allocation models of the different slices have converged faster than the baseline unconstrained IG post-hoc case. Here, the optimizer of EFL considers the relationships between the objectives and constraints of the two-player optimization problem, leading to improved performance compared to the uncon. IG post-hoc one, which accounts for only the objective function during optimization.
    \item \textbf{Sensitivity analysis:} To analyze our proposed model's sensitivity, we choose the recall metric, which is the rate of actual positive values for measuring the performance of our binary classification model. From Fig. 4, we can observe that the  recall score of the proposed one for all slices is in close proximity to the target threshold $\gamma$ (i.e., around $0.88\%$), which is an acceptable value for operators and slices' tenants.
    \item \textbf{Trustfulness:} In Fig. 5-(a), we observe the effect of changing the value top $p$\% on the log-odds, considering proposed model for all slices. Also we present a comparative analysis of log-odds score in Fig.5-(b) for both cases which proof the superiority of the proposed constrained EFL model. So, the statistics of the log-odds score give us an approximate idea of our model's reliability and trustworthiness. It shows that the log-odds score is decreasing with respect to the top $p$\% value, which conveys that our model is explainable and trustworthy in the training phase. 
\end{itemize}

\begin{figure}
\centering
     \subfloat[log-odds Score vs. top p $\%$ ]{\includegraphics[width=0.40\textwidth]{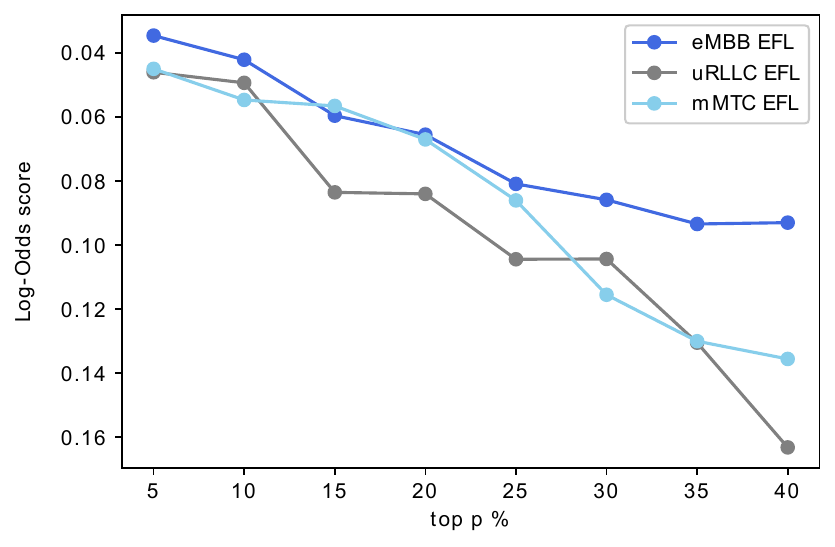}}\hspace{1cm}
     \subfloat[log-odds Score]{
     \includegraphics[width=0.40\textwidth]{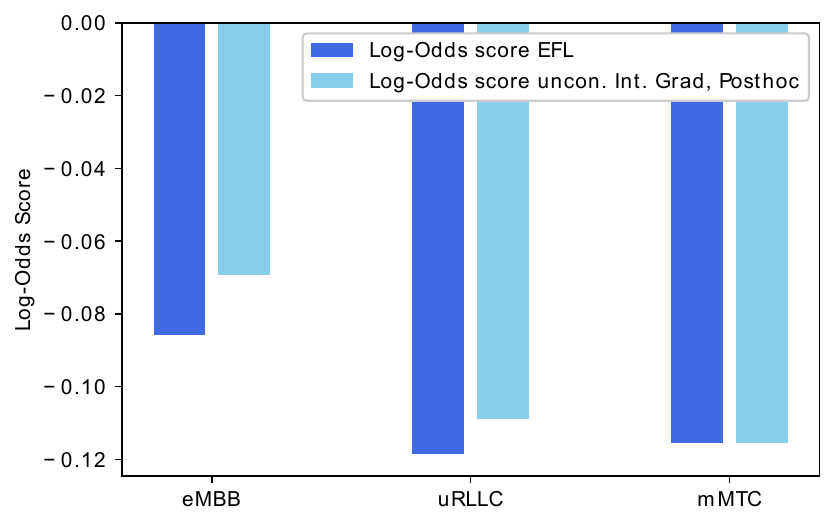}}
\caption{ Analysis of log-odds score with Lower bound of Recall score, $\alpha = [0.9, 0.95, 0.95]$ and Upper bound of log-odds score, $\beta = [-0.01, -0.01 , -0.01]$}
\vspace{-8mm}
\end{figure}
Furthermore, in Fig. 6, the correlation heatmaps of the proposed XAI method of the eMBB slice has presented for further analysis. It helps us visualize the strength of relationships between different variables and, in our case, identify which feature variation impacts the most for SLA variation. To plot correlation matrix heatmap, we consider one matrix, $\mathbf{R}_{k,n}$ = [$\mathbf{a}_{k,n},\hat{\mathbf{y}}_{k,n},\mathbf{y}_{k,n}$], where, $\mathbf{a}_{k,n}$ is the attribution score of features variable with dimensions $D_{k,n} \times Q$ and $\hat{\mathbf{z}}_{k,n}$ is the predicted output variable with dimensions $D_{k,n} \times 1$ and $\mathbf{y}_{k,n}$ is the true predicted value with dimensions $D_{k,n} \times 1$.
From the heatmap we see that the third feature, which is the channel quality, has the most impact on the recall value. If the third feature increases, the recall value will increase and vice versa.

\section{Conclusion}
This paper has presented a novel closed-loop explainable federated learning (EFL) approach to achieve transparent zero-touch service management of 6G network slices at RAN in a non-IID setup. We have jointly considered explainability and sensitivity metrics as constraints in the traffic drop prediction task, which we have solved using a proxy-Lagrangian two-player game strategy. From the results, we conclude that the proposed EFL scheme is reliable and trustful compared to state-of-the-art unconstrained post-hoc FL. Finally, the heatmaps of the attributions correlation matrix are presented to showcase the features whose variation influence more the traffic drop.

\section{Acknowledgment}
This work has been supported in part by the projects SEMANTIC (861165), 6G-BRICKS (101096954) HORIZON-JU-SNS-2022 and ADROIT6G (101095363) HORIZON-JU-SNS-2022.

\begin{figure}
\centering
\includegraphics[width=0.45\textwidth]{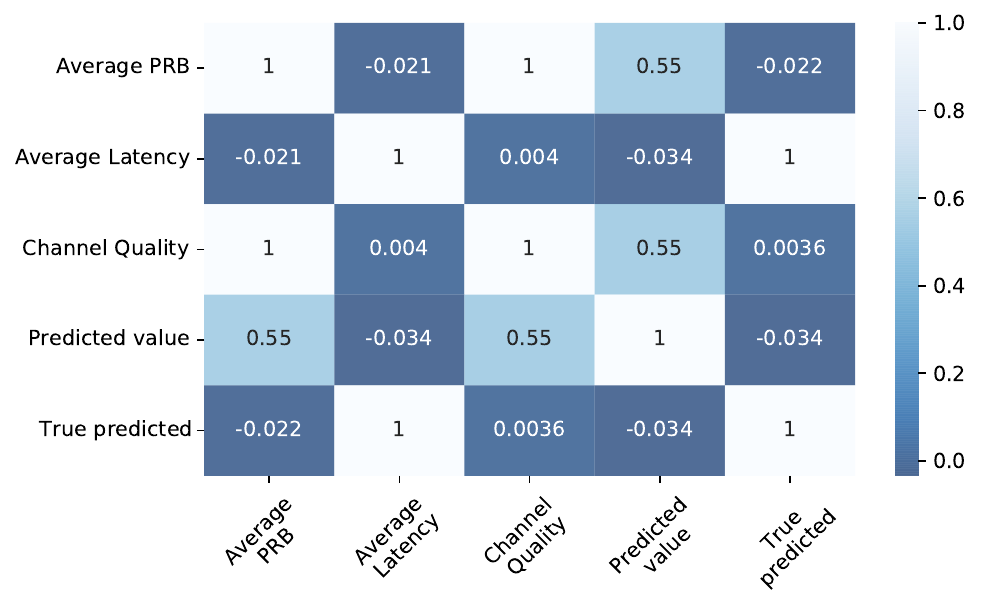}
\vspace{-3mm}
\caption{Correlation heatmap of eMBB slices based  on attribution scores of features generated by XAI.}
\vspace{-6mm}
\end{figure}
\vspace{1mm}
\bibliographystyle{IEEEtran}
\bibliography{myBibliographyFile}

\balance

\end{document}